\documentclass[conference]{IEEEtran}
\IEEEoverridecommandlockouts
\usepackage{url}
\usepackage{cite}
\usepackage{amsmath,amssymb,amsfonts}
\usepackage{algorithmic}
\usepackage{graphicx}
\usepackage{textcomp}
\usepackage{xcolor}
\usepackage[all]{nowidow}
\usepackage{multirow}
\usepackage{balance}
\usepackage[caption=false]{subfig}

\expandafter\def\expandafter\normalsize\expandafter{%
  \normalsize
  \setlength\belowdisplayskip{4ex}
  \setlength\belowdisplayshortskip{4ex}
}

\def\BibTeX{{\rm B\kern-.05em{\sc i\kern-.025em b}\kern-.08em
    T\kern-.1667em\lower.7ex\hbox{E}\kern-.125emX}}
\begin{document}

\title{QoS-Aware Resource Placement for LEO Satellite Edge Computing\\
    \thanks{Funded by the Deutsche Forschungsgemeinschaft (DFG, German Research Foundation) -- 415899119.}
}

\author{\IEEEauthorblockN{Tobias Pfandzelter, David Bermbach}
    \IEEEauthorblockA{\textit{Technische Universit\"at Berlin \& Einstein Center Digital Future}\\
        \textit{Mobile Cloud Computing Research Group} \\
        \{tp,db\}@mcc.tu-berlin.de}
}

\maketitle

\begin{abstract}
    With the advent of large LEO satellite communication networks to provide global broadband Internet access, interest in providing edge computing resources within LEO networks has emerged.
    The \emph{LEO Edge} promises low-latency, high-bandwidth access to compute and storage resources for a global base of clients and IoT devices regardless of their geographical location.

    Current proposals assume compute resources or service replicas at every LEO satellite, which requires high upfront investments and can lead to over-provisioning.
    To implement and use the LEO Edge efficiently, methods for server and service placement are required that help select an optimal subset of satellites as server or service replica locations.
    In this paper, we show how the existing research on resource placement on a 2D torus can be applied to this problem by leveraging the unique topology of LEO satellite networks.
    Further, we extend the existing discrete resource placement methods to allow placement with QoS constraints.
    In simulation of proposed LEO satellite communication networks, we show how QoS depends on orbital parameters and that our proposed method can take these effects into account where the existing approach cannot.
\end{abstract}


\section{Introduction}
\label{sec:introduction}

Technological advances such as free-space laser links, phased-array antennas, and re-usable rockets have enabled large low-Earth orbit (LEO) satellite constellations comprising thousands of satellites that provide global Internet access.
Built by private companies such as SpaceX\footnote{https://www.starlink.com}, Telesat\footnote{https://www.telesat.com}, or OneWeb\footnote{https://www.oneweb.world}, these networks promise low-latency, high-bandwidth connectivity anywhere on Earth, especially where terrestrial fiber is unavailable~\cite{Klenze2018-og,Handley2018-ay,Bhattacherjee2018-vc,Handley2019-ce,Del_Portillo2019-al}.

In light of these developments, recent proposals have explored the possibilities of bringing edge computing to LEO networks, the \emph{LEO Edge}~\cite{Zhang2019-ew,Bhattacherjee2020-kr,Bhosale2020-aa,pfandzelter2020edge,Pfandzelter2021-dp,wangtiansuan}.
Deploying servers and services to LEO satellites can provide significant quality-of-service (QoS) improvements to LEO network clients.
Yet equipping each of the thousands of satellites in a constellation with compute resources and running a replica of the respective edge service on it requires significant upfront investments and can result in costly over-provisioning.

Instead, we propose to use only a subset of satellites in the network as satellite servers or replica locations, such that all other network nodes can reach these resources with a specific QoS level, the service level objective (SLO).
In this paper, we propose an algorithmic approach to LEO Edge resource placement with respect to a network distance SLO.
To this end, we make the following contributions:

\begin{itemize}
    \item We show how LEO satellite networks can be modeled as 2D tori and introduce distance metrics resulting from orbital mechanics (Section~\ref{sec:torus}).
    \item We state the selection of satellites for server placement on the LEO network with QoS constraints as a distance\nobreakdash-$d$ resource placement problem on a 2D torus with edge weights and propose algorithms that solve this problem (Section~\ref{sec:solution}).
    \item We evaluate our proposed solution in simulations of proposed LEO satellite constellations (Section~\ref{sec:evaluation}).
    \item We discuss our work critically, show limitations of our approach, and derive avenues for future work (Section~\ref{sec:discussion}).
\end{itemize}

Additionally, we give an overview of LEO satellite constellations and the LEO Edge in Section~\ref{sec:background} and discuss existing research in this area in Section~\ref{sec:relwork}.

\section{LEO Satellite Networks \& the LEO Edge}
\label{sec:background}

In this section, we give an overview of LEO satellite networks and introduce the concept of LEO Edge computing.

\subsection{Large LEO Satellite Networks}

\begin{figure}
    \centering
    \includegraphics[width=0.9\linewidth]{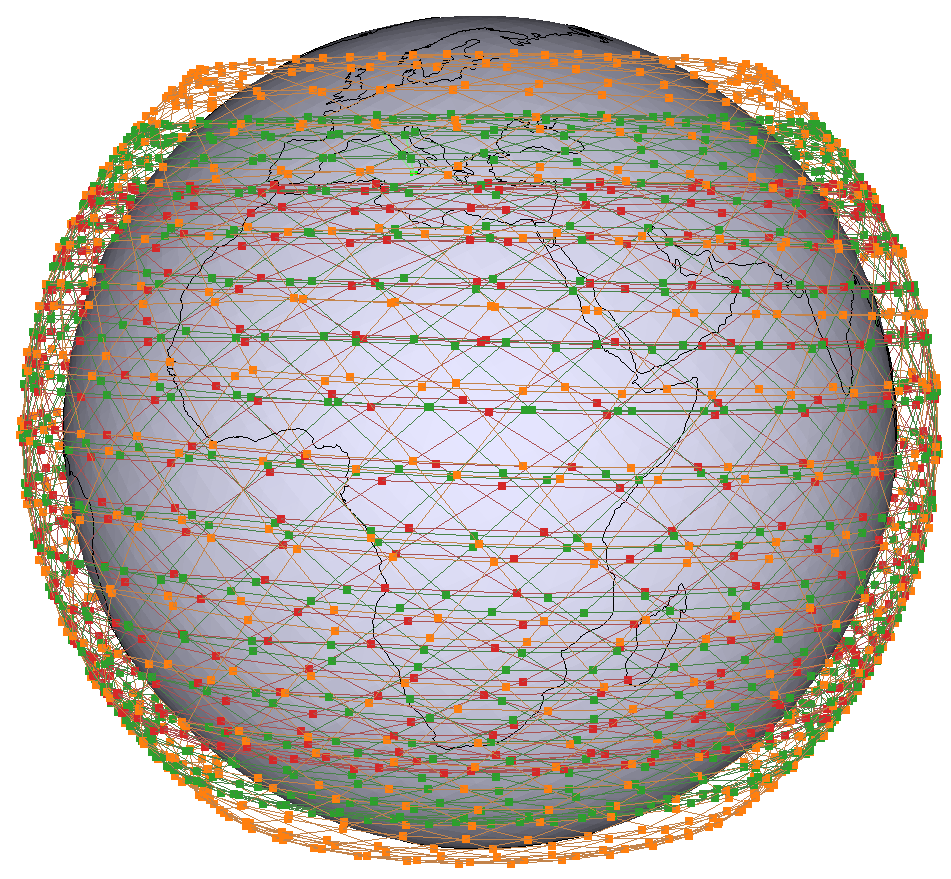}
    \caption{The proposed Kuiper constellation comprises three shells: 1,156 satellites at 630km/51.9° (orange), 784 at 590km/33° (red), 1,296 at 610km/42° (green)~\cite{Kassing2020-yc}.}
    \label{fig:kuiper}
\end{figure}

Satellite-based Internet access has traditionally used satellites in geostationary orbits at altitudes in excess of 35,000km, leading to significant transmission delays that are only feasible where no alternative connection technologies are available~\cite{Clarke1945-qb}.
In recent years, a new class of satellite communication networks has emerged as a result of technological advancements, with private aerospace companies such as SpaceX, Telesat, OneWeb, or Amazon Kuiper planning or already deploying networks of thousands of satellites in LEO, below altitudes of 2,000km~\cite{Handley2018-ay,Bhattacherjee2018-vc,Handley2019-ce,Del_Portillo2019-al}.

A complete network comprises multiple shells in which satellites are evenly spaced on an orbital plane, and multiple such orbital planes are spaced evenly along the equator~\cite{walker1984satellite,408677}.
All satellites within a shell share orbital parameters such as their altitude and their inclination, which describes the orbital plane's angle to the equator.
Figure~\ref{fig:kuiper} shows the proposed Amazon Kuiper network.
The first shell of this network comprises 1,156 satellites at an altitude of 630km and an inclination of 51.9°, with 34 satellites in each of the 34 orbital planes~\cite{Kassing2020-yc}.
As a result of the low altitude, satellites have low orbital periods, e.g., 97 minutes at 630km, at a speed of 27,150km/h.
In addition, the Earth revolves below the satellite constellations.
Consequently, ground stations frequently need to reconnect to their nearest satellite and, as a difference to geostationary orbits, satellites continuously cover new geographical areas.

In addition to relaying radio signals from ground stations, satellites within a shell connect to each other using inter-satellite laser links (ISL).
This follows the \emph{+GRID} pattern, where each satellite connects to its successor and predecessor within its orbital plane as well as the nearest satellite from each adjacent plane.
This makes point-to-point connections between any two ground stations on Earth possible.
As lasers in the vacuum of space can benefit from a 47\% faster light propagation than in fiber, satellite networks can offer a significantly reduced network delay compared to terrestrial networks~\cite{Handley2018-ay,Bhattacherjee2019-jz}.

\subsection{LEO Edge Computing}

In edge computing, compute and storage resources are embedded within the network and close to consumers to offer low access latency, increased throughput, increased privacy, and reduced network costs~\cite{paper_bonomi_fog,Zhang2015-cb,paper_bermbach_fog_vision,Grambow2018-um}.
For LEO satellite networks, the network edge is the satellite constellation itself, as satellites communicate directly with user equipment, i.e., the ground stations.
With the LEO Edge, researchers have thus proposed to add compute resources to LEO satellites to build edge applications such as CDNs or IoT data preprocessors.
As uplinks for multiple ground stations converge in a single satellite, the LEO Edge clients can share these resources efficiently~\cite{Bhattacherjee2020-kr,pfandzelter2020edge,Pfandzelter2021-dp,wangtiansuan,Wu2016-rm}.

Several challenges still lie ahead before any LEO Edge infrastructure is made publicly available:
Current service management paradigms will have to be adapted for the high degree of mobility and geographical distribution of LEO satellite constellations~\cite{Bhosale2020-aa,Pfandzelter2021-dp}.
Then, engineering challenges such as the effects of radiation on compute hardware or heat dissipation will need to be addressed~\cite{Nedeau1998-fj,Koontz2018-pb,noauthor_2021-jm}.
LEO Edge infrastructure will also require significant upfront investments, and it is unlikely that operators equip thousands of satellites with the necessary hardware immediately.
Instead, we propose to add such hardware only to a carefully chosen subset of satellites to limit upfront costs while still achieving set QoS targets.
Specifically, we focus on network latency in this paper, which is a direct result of ISL distances, as we will show.
Resource placement has limited influence on other QoS factors and dimensions such as computation delay or data consistency.

\section{Modeling LEO Satellite Networks as 2D Tori}
\label{sec:torus}

While terrestrial edge networks are flexible and resources can be allocated wherever there is demand from clients, the LEO Edge must obey the laws of orbital mechanics.
As LEO communication satellites are highly mobile and constantly serve new areas of clients, we cannot simply allocate compute resources and services where they are actually needed but must place them for global coverage.

\begin{figure}
    \centering
    \includegraphics[width=\linewidth]{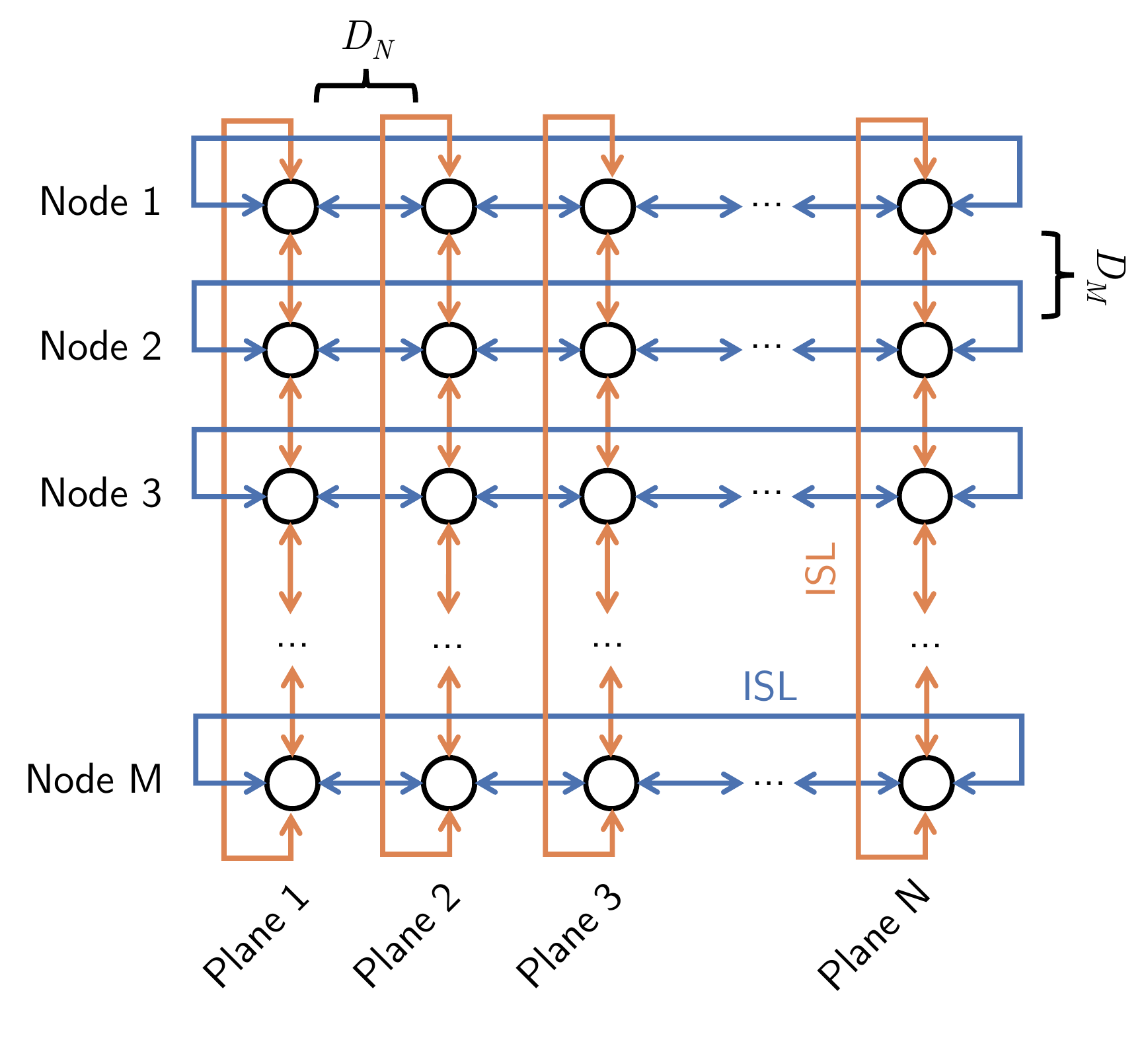}
    \caption{The satellite network with its ISLs can be modeled as an $N \times \ M$ 2D torus, where $N$ is the number of orbital planes and $M$ is the number of satellites per plane. The length of a hop is constant within an orbital plane but can vary between adjacent nodes over the satellite's orbital period.}
    \label{fig:torus}
\end{figure}

\begin{figure}
    \centering
    \includegraphics[width=0.9\linewidth]{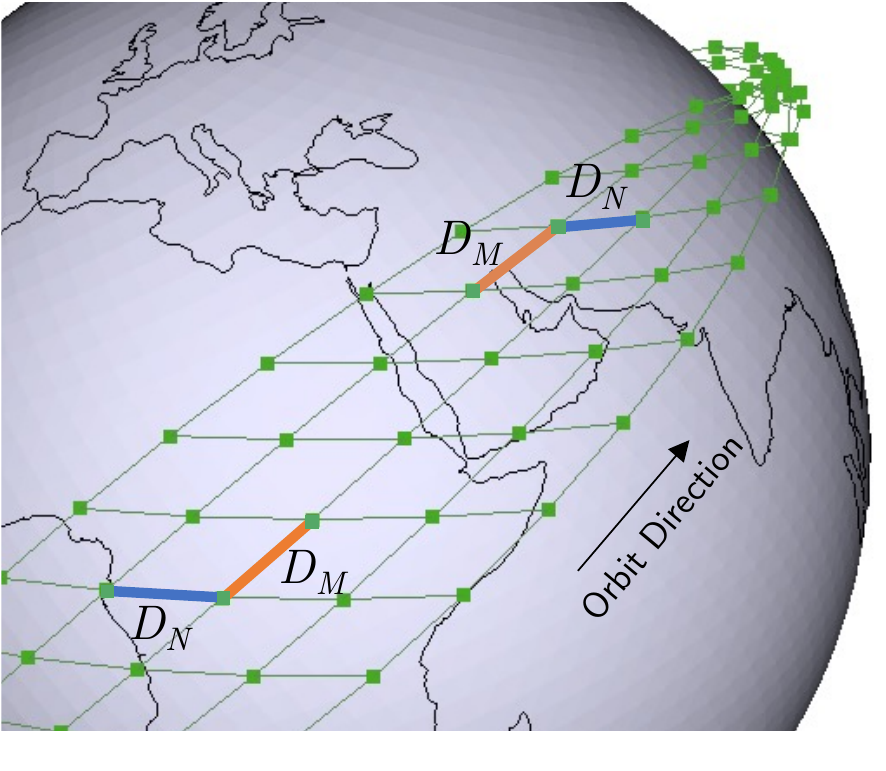}
    \caption{As satellites within a plane are evenly spaced along their orbit, $D_{M}$ remains constant. $D_{N}$ is larger near the equator as satellites in adjacent planes are closer near the poles. Adjacent, connected planes all orbit in the same direction.}
    \label{fig:distances}
\end{figure}

To achieve such a placement, we first create a model of the LEO satellite network that reflects the length of ISLs over time.
Despite the high mobility, the actual topology of the network does not change as ISL pairs remain constant as a result of the +GRID configuration.
With each satellite having four ISL neighbors, each shell is an $N \times M$ 2D torus, where $N$ is the number of orbital planes and $M$ is the number of satellites per plane, as shown in Figure~\ref{fig:torus}~\cite{Sun2002-fl,Ji2021-lq}.
A 2D torus is similar to a mesh with the addition of ``wraparound'' edges that connect first and last nodes vertically and horizontally.
In the satellite network, these are the links between the first and last node of an orbital plane, and the links between first and last orbital plane of a shell, respectively.
Note that with the +GRID configuration, there are no links between different shells of a complete LEO constellations.
Each shell can thus be modelled individually.

Figure~\ref{fig:distances} shows how ISL distances for neighboring satellites change.
Satellites within a plane are evenly spaced along their orbit.
Their distance, that we denote as $D_{M}$, remains constant and is given by:

\begin{equation*}
    D_{M} = (r_{E} + h) \sqrt{2(1 - \cos(\frac{2\pi}{M}))}
\end{equation*}

, where $r_{E}$ is the radius of the Earth and $h$ is the orbital altitude.
Please, note that we here assume a perfectly spherical Earth, which is inaccurate but simplifies the model.
In our simulations, we use a more accurate Earth model to evaluate whether our algorithms can be transferred to the real world.

The distance of satellites in adjacent planes, that we denote as $D_{N}$, follows an ellipse and varies over the satellites' orbital periods $T$.
The planes are closer near the poles, dependent on the orbit inclination $i$.
At a time in a satellite's orbital period $t \in[0, T]$:

\begin{equation*}
    \resizebox{\linewidth}{!}{
        $D_{N}(t) = (r_{E} + h) \sqrt{2(1 - \cos(\frac{2\pi}{N}))} \sqrt{  \cos^{2}(2\pi\frac{t}{T}) + \cos^{2}(i) \sin^{2}(2\pi\frac{t}{T}) }$
    }
\end{equation*}

We assume that $t = 0$ is the point at which the satellite intersects the equatorial plane of the Earth (\emph{ascending node}).
The \emph{maximum} distance between two satellites in adjacent planes, that we denote as ${D_{N}^{\text{max}}}$, is thus reached at the equator ($t\in\{0, \frac{T}{2}, T\}$):

\begin{equation*}
    {D_{N}^{\text{max}}} = \max\limits_{t \in[0, T]} D_{N} = (r_{E} + h) \sqrt{2(1 - \cos(\frac{2\pi}{N}))}
\end{equation*}

As a result, note that ${D_{N}^{\text{max}}} = D_{M}$  for quadratic constellations, i.e., where $N = M$.

Further, the \emph{mean} distance between two satellites in adjacent planes $\bar{D}_{N}$ is given by:

\begin{align*}
    \bar{D}_{N} & = \frac{1}{T} \int_0^T D_{N}(t) dx                                              \\
                & = \frac{2}{\pi} (r_{E} + h) \sqrt{2(1 - \cos(\frac{2\pi}{N}))} E(1-\cos^{2}(i))
\end{align*}

, where $E$ is the complete elliptic integral of the second kind.
To model the length of ISLs in the satellite constellation, we can thus use both maximum path distances, which provide a hard limit that cannot be exceeded, and the mean path distance that is the mean network distance  over time.
For laser ISLs in a vacuum, these spatial distance metrics can be converted into communication delays by dividing by the speed of light in a vacuum $c$.

\section{Resource Node Placement With QoS Constraints}
\label{sec:solution}

Given a QoS constraint $\mathbb{D}_{\text{SLO}}$, our goal is to find a subset of satellite nodes $R \subseteq S$ (\emph{resource nodes}), so that for every node $s \in S$, the distance of the shortest path $D_{i,j}, i,j \in S$ to at least one selected resource node $r \in R$ (\emph{resource distance}) is less than or equal to $\mathbb{D}_{\text{SLO}}$:

\begin{equation*}
    D_{s,r} \le \mathbb{D}_{\text{SLO}}, \forall s\in S.\exists r\in R
\end{equation*}

To save costs, we want to find a solution that meets these constraints while using the minimum number of resource nodes.

\begin{figure*}
    \centering
    \subfloat[The original torus has dimensions $N = 5$ and $M = 3$, and hop lengths $D_{N} = 0.8$ and $D_{M} = 1.3$.\label{fig:algorithm:1}]{
        \includegraphics[width=0.2\textwidth]{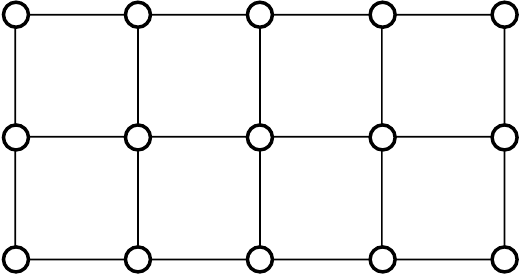}
    }
    \hfill
    \subfloat[The modified torus has dimensions $N = 5$ and $M' = 5$, and equal hop lengths.\label{fig:algorithm:2}]{
        \includegraphics[width=0.2\textwidth]{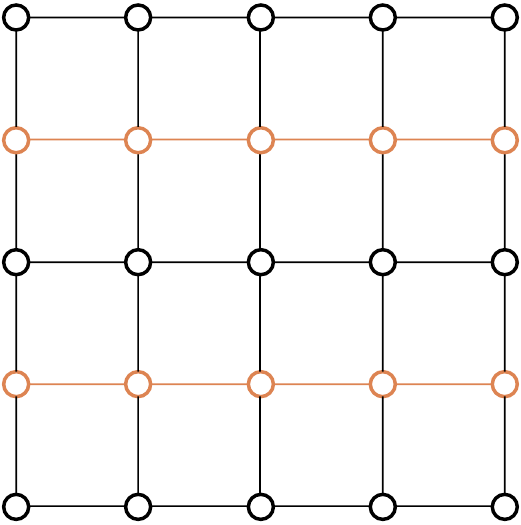}
    }
    \hfill
    \subfloat[A $1$-hop placement is performed.\label{fig:algorithm:3}]{
        \includegraphics[width=0.2\textwidth]{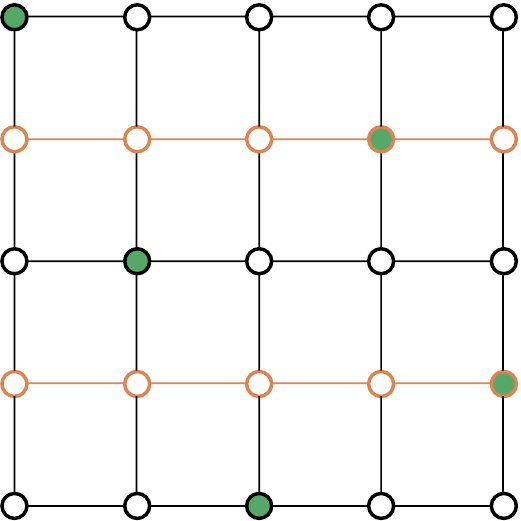}
    }
    \hfill
    \subfloat[The placement is transferred to the original torus by multiplying $y$-coordinates with $\frac{M}{M'}$.\label{fig:algorithm:4}]{
        \includegraphics[width=0.2\textwidth]{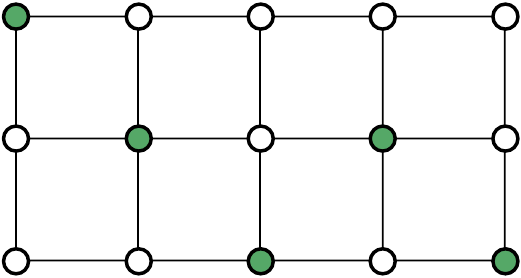}
    }
    \caption{Placing resource nodes with distance $d = 1.4$ on a $5 \times 3$ torus with $D_{N} = 0.8$ and $D_{M} = 1.3$ (\ref{fig:algorithm:1}, wraparound edges not shown): the torus is extended to $5 \times \lceil 3\frac{D_{M}}{D_{N}}\rceil$ (\ref{fig:algorithm:2}), then a $\lfloor\frac{d}{D_{N}}\rfloor$-hops placement is performed (\ref{fig:algorithm:3}). Finally, $y$-coordinates are transferred by multiplying with $\frac{3}{5}$, yielding the resource placement on the original torus (\ref{fig:algorithm:4}).}
    \label{fig:algorithm}
\end{figure*}

\subsection{$d$-Hops Placement on Unweighted Torus Networks}

When considering the 2D torus network as an unweighted graph and hops on that graph as a distance metric, we get the $d$-hops problem~\cite{Bae1997-gt}, where ${d \in \mathbb{N}}$.
A unique, regular, and perfect $d$-hops placement has every non-resource node at a distance of $d$ or less to \emph{exactly one} resource node.
This is possible if and only if $N$ and $M$ are multiples of $k$, where ${k = 2d^2 + 2d + 1}$~\cite{Albdaiwi2004-fb}.
A resource node placement for the ${k \times k}$ torus is obtained by ${[i, 2d^2i] \pmod{k}, i = 0, 1, \dotsc, (k-1)}$.
The original torus can then be tiled with the ${k \times k}$ torus~\cite{Bae1997-gt}.
For 2D tori where these constraints do not hold, i.e., of dimensions ${N = pk + r}$ and ${M = qk + s}$ with ${p, q, s, r \in \mathbb{N}}$ and ${0 < s, r < k}$, Bae and Bose~\cite{Bae1996-ys} propose building ${(p + 1) \times (q + 1)}$ blocks of the ${k \times k}$ torus.
For these, it is possible to construct a perfect $d$-hops placement.
Finally, ${k - r}$ rows and ${k - s}$ columns are removed before adding additional resource nodes for nodes of a distance more than $d$ hops.
While this is not a \emph{perfect} placement, as some nodes are of a distance of $d$ hops or less to \emph{more than one} resource node, it leads to a solution with the smallest possible number of nodes.

\subsection{Real $d$-Distance Placement}

The existing method for $d$-hops placement cannot be applied directly to a torus where vertical and horizontal distances are real and not equal.
Instead, we propose normalizing these distances first by introducing virtual nodes.
We assume an $N \times M$ torus with horizontal hop length $D_{N}$ and vertical hop length $D_{M}$, where $D_{N}, D_{M} \in \mathbb{R}_{>0}$ and $D_{N} \le D_{M}$.
Given a target distance $d \in \mathbb{R}_{>0}$ with $D_{N} \le d$, construct the $N \times M'$ torus with $M' = \lceil M\frac{D_{M}}{D_{N}}\rceil$.
On this torus we perform a $\lfloor\frac{d}{D_{N}}\rfloor$-hops placement following the method for integer $d$-hops placement.
Finally, we transfer the placement to the $N \times M$ torus by multiplying $y$-coordinates with $\frac{M}{M'}$.
All nodes then have a resource node at distance $d + \epsilon$ or less, where $\epsilon$ is the error introduced by rounding.
Figure~\ref{fig:algorithm} illustrates this process.

Note that when $D_{M} < D_{N}$, we can instead apply the method on the $M \times N$ torus.
When $D_{N} \le d < D_{M}$ or $D_{M} \le d < D_{N}$, tile the $N \times M$ torus with an $N \times 1$ or $1 \times M$ torus, respectively.
Where $d < D_{N}, D_{M}$, every node must be a resource node.

\section{Evaluation}
\label{sec:evaluation}

To evaluate our approaches, we simulate LEO satellite networks with an extended version of the \emph{SILLEO-SCNS} LEO satellite network simulator~\cite{Kempton2021-lw}.
Our extension supports the more accurate \emph{SGP4} simplified perturbations models and \emph{WGS84} world geodetic system, and we make it available as open-source.\footnote{https://github.com/pfandzelter/optimal-leo-placement}

\subsection{Parameters}

We consider shells of the phase~\textrm{I} Starlink and Kuiper constellations.
Orbital parameters of these shells are given in Table~\ref{tab:orbital}.
For each of these shells, we construct resource placements with SLOs of different metrics: $1$-hop and $4$-hops, and maximum and mean distances of $10\text{ms} \times c$ ($2997.92\text{km}$) and $100\text{ms} \times c$ ($29979.24\text{km}$).
We assign a fixed resource satellite to each satellite and observe simulated resource distances over time.
Our simulation spans one day (86,400 seconds) at a one-second simulation interval.

\begin{table}
    \renewcommand{\arraystretch}{1.3}
    \caption{Orbital Parameters for Shells Considered in Our Simulation~\cite{Kassing2020-yc}}
    \label{tab:orbital}
    \centering
    \resizebox{\columnwidth}{!}{
        \begin{tabular}{c|c|r|r|r|r}
            \hline
            \textbf{Constellation}                     & \textbf{Shell} & \multicolumn{1}{c|}{\textbf{\begin{tabular}[c]{@{}c@{}}Planes\\ ($N$)\end{tabular}}} & \multicolumn{1}{c|}{\textbf{\begin{tabular}[c]{@{}c@{}}Satellites/Plane\\ ($M$)\end{tabular}}} & \multicolumn{1}{c|}{\textbf{\begin{tabular}[c]{@{}c@{}}Altitude\\ in km\end{tabular}}} & \multicolumn{1}{c}{\textbf{Inclination}} \\ \hline\hline
            \multirow{2}{*}{\begin{tabular}[c]{@{}c@{}}SpaceX\\ Starlink\end{tabular}} & A              & 72                                                      & 22                                                      & 550                                                     & 53.0°                                    \\ \cline{2-6}
                                                       & B              & 5                                                       & 75                                                      & 1,275                                                   & 81.0°                                    \\ \hline
            \multirow{2}{*}{\begin{tabular}[c]{@{}c@{}}Amazon\\ Kuiper\end{tabular}} & A              & 34                                                      & 34                                                      & 630                                                     & 51.9°                                    \\ \cline{2-6}
                                                       & B              & 28                                                      & 28                                                      & 590                                                     & 33.0°                                    \\
            \hline
        \end{tabular}
    }
\end{table}

\subsection{Placements}

\begin{table}
    \renewcommand{\arraystretch}{1.3}
    \caption{Number of Resource Nodes Required for Placements on Shells With Different QoS Targets}
    \label{tab:placements}
    \centering
    \resizebox{\columnwidth}{!}{
        \begin{tabular}{c|r|rrrrrr}
            \hline
            \multirow{2}{*}{\textbf{Shell}} & \multicolumn{1}{c|}{\multirow{2}{*}{\textbf{\# Nodes}}} & \multicolumn{6}{c}{\textbf{\# Resource Nodes}}                                                                                                                                                                                                                                  \\ \cline{3-8}
                                            & \multicolumn{1}{c|}{}                                   & \multicolumn{1}{c|}{1-Hop}                     & \multicolumn{1}{c|}{4-Hops} & \multicolumn{1}{c|}{\begin{tabular}[c]{@{}c@{}}Mean\\ $10\text{ms}$\end{tabular}} & \multicolumn{1}{c|}{\begin{tabular}[c]{@{}c@{}}Max.\\ $10\text{ms}$\end{tabular}} & \multicolumn{1}{c|}{\begin{tabular}[c]{@{}c@{}}Mean\\ $100\text{ms}$\end{tabular}} & \multicolumn{1}{c}{\begin{tabular}[c]{@{}c@{}}Max.\\ $100\text{ms}$\end{tabular}} \\ \hline\hline
            Starlink A                      & 1,584                                                   & \multicolumn{1}{r|}{354}                       & \multicolumn{1}{r|}{55}     & \multicolumn{1}{r|}{94}                        & \multicolumn{1}{r|}{135}                       & \multicolumn{1}{r|}{2}                         & 2                                             \\ \hline
            Starlink B                      & 375                                                     & \multicolumn{1}{r|}{75}                        & \multicolumn{1}{r|}{17}     & \multicolumn{1}{r|}{45}                        & \multicolumn{1}{r|}{45}                        & \multicolumn{1}{r|}{2}                         & 2                                             \\ \hline
            Kuiper A                        & 1,156                                                   & \multicolumn{1}{r|}{245}                       & \multicolumn{1}{r|}{41}     & \multicolumn{1}{r|}{128}                       & \multicolumn{1}{r|}{106}                       & \multicolumn{1}{r|}{2}                         & 2                                             \\ \hline
            Kuiper B                        & 784                                                     & \multicolumn{1}{r|}{178}                       & \multicolumn{1}{r|}{25}     & \multicolumn{1}{r|}{80}                        & \multicolumn{1}{r|}{178}                       & \multicolumn{1}{r|}{2}                         & 2                                             \\ \hline
        \end{tabular}
    }
\end{table}

\begin{figure*}
    \centering
    \subfloat[Mean Resource Distance in km\label{fig:results:mean}]{
        \includegraphics[width=0.48\linewidth]{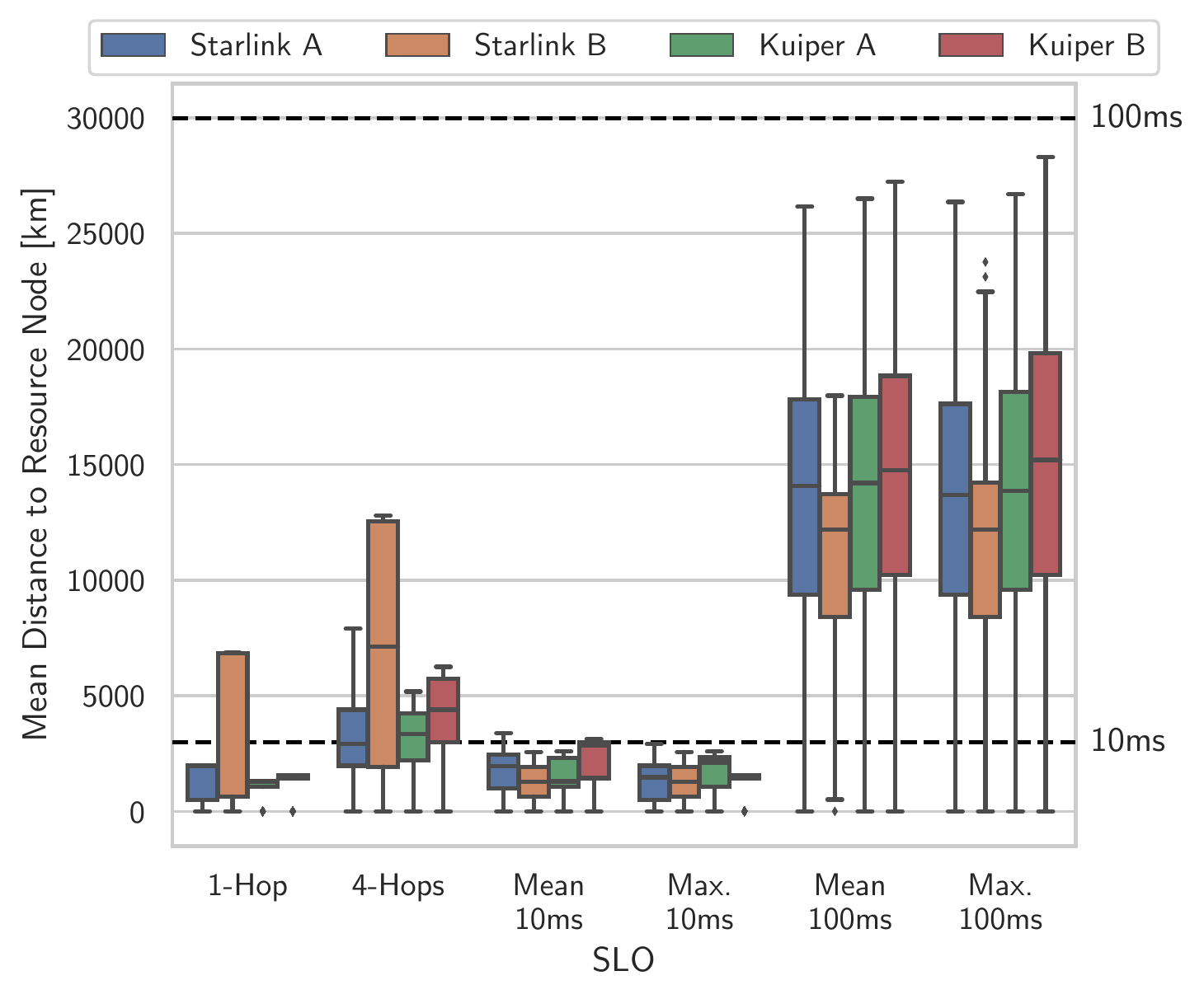}
    }
    \hfill
    \subfloat[Maximum Resource Distance in km\label{fig:results:max}]{
        \includegraphics[width=0.48\linewidth]{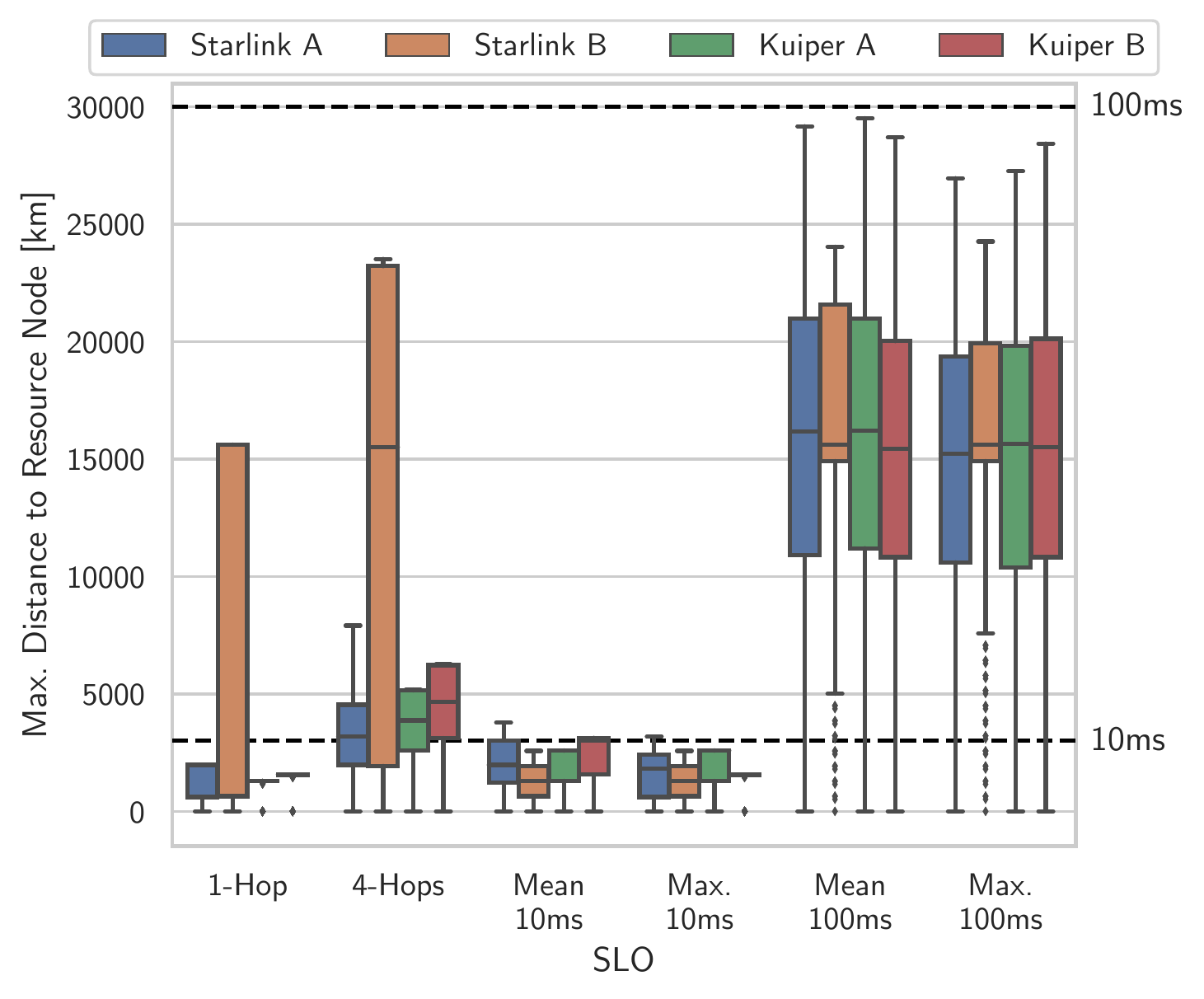}
    }
    \caption{Aggregated simulated distance to assigned resource node in different shells with different QoS targets over the course of one day.}
    \label{fig:results}
\end{figure*}
Table~\ref{tab:placements} shows the results of placement under different QoS constraints on the shells we consider.
The number of resource nodes required for $d$-hop placement scales linearly with the total number of nodes in the shell, as the actual distance between nodes is not factored in.
For all shells, two nodes are sufficient to achieve a mean and maximum SLO of $100\text{ms}$, a result of the low altitude of the LEO shells.
We expect the maximum $10\text{ms}$ SLO to require more resource nodes than mean $10\text{ms}$.
In case of the Starlink~B shell, the low number of planes leads to a high distance between planes.
As a result, each plane is tiled individually and the difference between maximum and mean distances is not taken into account.
The placement for the Kuiper~A shell requires more resource nodes for the mean $10\text{ms}$ SLO than for maximum $10\text{ms}$.
We attribute this effect to inaccuracies introduced by rounding, amplified by the fact that $D_{M} = D_{N}^{\text{max}}$ as $N = M$.

\subsection{Results}

We show the mean and maximum distances between all satellite and their assigned resource nodes in Figures~\ref{fig:results:mean} and~\ref{fig:results:max}.
We see how the resource node distance in discrete $d$-hops placements depends on the orbital parameters of the shell:
With a higher altitude and fewer satellites, the lengths of hops, i.e., ISL distances, increase.
Notably, the Starlink~B shell has only five orbital planes, which results in a large maximum distance between satellites in adjacent planes.
The high orbital inclination of 81.0° leads to a decreased mean distance.
These results show that $d$-hops placement is insufficient to guarantee SLOs.

Real $d$-distance placement, on the other hand, performs as expected:
Regardless of orbital parameters, the $10\text{ms}$ and $100\text{ms}$ SLOs are adhered to with their respective placements.
Notably, for the Starlink~B shell we achieve better QoS with maximum and mean $10\text{ms}$ placement than $1$-hop placement while also requiring less resource nodes.

\section{Discussion \& Future Work}
\label{sec:discussion}

We have shown how the $d$-hops placement algorithm for 2D tori can be extended for QoS-aware resource placement in satellite networks.
Nevertheless, some open questions remain that warrant further research.

\subsection{Application for Service Placement}

Unlike static compute resource placement, service placement, e.g., placing a database within a given distance of clients on the satellite network, can leverage service migration.
As edge services tend to have local relevance, e.g., a database with information on users in a certain country, service migration can be used to keep state stationary while the physical satellites move~\cite{paper_bonomi_fog,paper_bermbach_fog_vision,pfandzelter2020edge}.
Depending on the use case, it might prove feasible to select initial service locations with our proposed method and continuously migrate the service as needed.
This \emph{virtual stationarity} keeps services' locations fixed in relation to Earth.

Beyond that, skews in service popularity, e.g., with more clients located in population centers, might mean that a uniformly distributed placement of services leads to over-provisioning in some areas and under-provisioning in others.
In such cases, a dynamic, demand-driven resource placement combined with service migration might be a better approach.

\subsection{Fault Tolerance \& Availability}

Our proposed method distributed resource nodes evenly across the satellite network, with exactly one resource node within the target distance of any node wherever possible.
The failure of a single resource node can thus break QoS guarantees for a number of non-resource nodes.
With the \emph{Spaceborne Computers} on board the International Space Station, HPE have shown that using commercial, off-the-shelf compute hardware is possible in LEO, yet satellite servers will be subject to frequent, intermittent failures, e.g., caused by single event upset as a result of galactic cosmic rays~\cite{Nedeau1998-fj,noauthor_2021-jm,noauthor_2019-ui,Koontz2018-pb}.
The impact of such failure can be alleviated by choosing stricter SLOs, at the cost of more resource nodes.
In commercial deployment, operators must thus weigh off the cost of resources and the potential impact and expected frequency of equipment failure.

\subsection{Incremental Deployment of Satellite Servers}

Due to their size, satellite constellations are built incrementally, with a handful of satellites launched at a time.
After their lifespan expires, they are de-orbited and can be replaced by newer models~\cite{Sheetz_undated-dm}.
A similar model could be economically feasible for satellite servers:
If the operator plans to test the demand for such infrastructure first, they may start with a lower QoS.
Here, our approach allows incremental planning of satellite server deployment with increasing QoS targets.

\subsection{Routing \& Link Congestion}

The larger the coverage area of a resource node, i.e., the lower the number of resource nodes in the network, the more messages a resource node will receive from non-resource nodes.
Consequently, message routing must be addressed, especially as there may be multiple paths of equal length between two nodes.
Additionally, the ISLs of the resource nodes will have more load than those of non-resource nodes.
To remedy link congestion and distribute load, routes for other traffic may need to be adapted to avoid resource nodes~\cite{Bose1995-cu,Azizoglu2000-lx}.

\subsection{Other QoS Dimensions \& Factors}

In this paper, we have focused on network latency as a QoS factor, which is a direct result of physical distances between satellites.
Beyond that, quality of the LEO Edge, whether regarding the infrastructure or a particular edge service, can also include general performance, availability, scalability, and others~\cite{book_cloud_service_benchmarking}.
Resource placement, however, has limited to no impact on those dimensions, with some exceptions as mentioned above.
Choosing, e.g., the amount of compute resources to allocate for each selected satellite server, influences computational delays, scalability, or throughput, but is an orthogonal challenge to that of selecting satellites for server placement.

\section{Related Work}
\label{sec:relwork}

Resource placement has received considerable research interest in the context of fog and edge computing~\cite{Tong2016-ke,Naas2017-ln,Bittencourt2017-rf,paper_hasenburg_fogexplorer_2018,Bermbach2020-hg,Bermbach2021-pd,Pfandzelter2021-zp}.
For example, Brogi et al.~present \emph{FogTorch}~\cite{Brogi2017-fk} and \emph{FogTorch}$\Pi$~\cite{Brogi2017-nl}, approaches to service placement in the edge-cloud continuum.
Such service placement must consider a number of restrictions such as SLOs, resource costs, service requirements, and network characteristics.
While important in their own right, such approaches are not applicable to resource placement in the LEO Edge if they cannot consider high satellite server mobility.
Further, Brogi et al.~show that this resource allocation problem is NP-hard on arbitrary topologies.
Leveraging the toroid unique toroid topology of LEO satellite communication networks, we can achieve a more efficient solution.

Sun and Modiano~\cite{Sun2002-fl} use these characteristics to investigate  network routing, capacity provisioning, and failure recovery in symmetric $N \times N$ toroid satellite networks, but do not consider resource placement.
Ji et al.~\cite{Ji2021-lq} use known $d$-hops placement techniques to optimize the network control structure of a LEO satellite communications network, yet their approach does not address physical ISL distances.

The existing approaches to $d$-hops placement, as discussed, e.g., by Livingston and Stout~\cite{Livingston1990-hq}, Bae and Bose~\cite{Bae1996-ys}, or Albdaiwi and Livingston~\cite{Albdaiwi2004-fb}, only consider discrete distances, as they are based on discrete Lee distance error-correcting codes.
The driving motivation for such approaches were multiprocessor computers arranged in a torus topology, with, e.g., Azeez et al.~\cite{Azeez2006-mq} developing a method for I/O node placement.
Bae and Bose~\cite{Bae1996-ys} also include a scheme for system reconfiguration in presence of node failures using spare nodes, which could also be applied to increase availability for the LEO Edge.

\section{Conclusion}
\label{sec:conclusion}

LEO satellite edge computing can bring the benefits of edge computing to a global, geo-distributed base of clients and IoT devices, enabling novel applications and services.
In this paper, we have motivated the need for resource placement for the LEO Edge, allowing an efficient implementation of satellite servers and edge services.
Specifically, we have shown how orbital parameters have a direct impact on network delays.
By modelling LEO satellite constellations as 2D tori, we were able to apply an existing resource placement method.
Further, we have extended this method to take SLOs into account and were thus able to achieve QoS-aware resource placements.
In simulation of real satellite constellations, we have shown that our proposed algorithm achieves QoS targets where the existing discrete method cannot.

\balance
\bibliographystyle{IEEEtran}
\bibliography{bibliography}

\end{document}